# AN APPROACH FOR CLASSIFICATION OF DYSFLUENT AND FLUENT SPEECH USING K-NN AND SVM

P.Mahesha[1] and D.S.Vinod[2]

[1]Department of Computer Science and Engineering, Sri Jayachamarajendra College of Engineering, Mysore, Karnataka, India
maheshsjce@yahoo.com
[2]Department of Information Science and Engineering, Sri Jayachamarajendra College of Engineering, Mysore, Karnataka, India
dsvinod@daad-alumni.de

## ABSTRACT

*This paper presents a new approach for classification of dysfluent and fluent speech using Mel-Frequency Cepstral Coefficient (MFCC). The speech is fluent when person's speech flows easily and smoothly. Sounds combine into syllable, syllables mix together into words and words link into sentences with little effort. When someone's speech is dyfluent, it is irregular and does not flow effortlessly. Therefore, a dysfluency is a break in the smooth, meaningful flow of speech. Stuttering is one such disorder in which the fluent flow of speech is disrupted by occurrences of dysfluencies such as repetitions, prolongations, interjections and so on. In this work we have considered three types of dysfluencies such as repetition, prolongation and interjection to characterize dysfluent speech. After obtaining dysfluent and fluent speech, the speech signals are analyzed in order to extract MFCC features. The k-Nearest Neighbour (k-NN) and Support Vector Machine (SVM) classifiers are used to classify the speech as dysfluent and fluent speech. The 80% of the data is used for training and 20% for testing. The average accuracy of 86.67% and 93.34% is obtained for dysfluent and fluent speech respectively.*

## 1. INTRODUCTION

Stuttering also known as dysphemia and stammering is a speech fluency disorder that affects the flow of speech. It is one of the serious problems in speech pathology and poorly understood disorder. Approximately about 1% of the population suffering from this disorder and has found to affect four times as many males as females [11, 5, 16, 3]. Stuttering is the subject of interest to researchers from various domains like speech physiology, pathology, psychology, acoustics and signal analysis. Therefore, this area is a multidisciplinary research field of science.

The speech fluency can be defined in terms of continuity, rate, co-articulation and effort. Continuity relates to the degree to which syllables and words are logically sequenced and also the presence or absence of pauses. If semantic units follow one another in a continual and logical flow of information, the speech is interpreted as fluent [4]. If there is a break in the smooth, meaningful flow of speech, then it is dysfluent speech. The types of dysfluency that characterize stuttering disorder are shown in Table 1 [6].

            23

International Journal of Computer Science, Engineering and Applications (IJCSEA) Vol.2, No.6, December 2012There are not many clear and quantifiable characteristic to distinguish the dysfluencies of dysfluent and fluent speakers. It was found from literature survey that sound or syllable repetitions, word repetitions and prolongation are sufficient to differentiate them [6, 12].

Table 1. Types of dysfluencies

| **Repetition** | Syllable repetition (The baby ate the s-s-soup). |
| --- | --- |
| | Whole word repetition (The baby-baby ate the soup) |
| | Phrase or sentence repetition (The baby-the baby ate the soup). |
| **Prolongation** | Syllable prolongation (The baaaby ate the soup). |
| **Interjection** | Common interjections are "um" and "uh" (The baby um ate the um soup). |
| **Pauses** | The [pause] baby ate the [pause] soup. Silent duration within speech considered fluent and considered as dysfluency, if they last more than 2 sec. |

There are number of diagnosis methods to evaluate stuttering. The stuttering assessment process is carried out by transcribing the recorded speech and locating the dysfluencies occurred and counting the number of occurrences. These types of stuttering assessments are based on the knowledge and experience of speech pathologist. The main drawbacks of making such assessment are time consuming, subjective, inconsistent and prone to error.

In this work, we are proposing an approach to classify dysfluent and fluent speech using MFCC feature extraction. In order to classify stuttered speech we have considered three types of dyfluencies such as repetition, prolongation and interjection.

## 2. SPEECH DATA

The speech samples are obtained from University College London Archive of Stuttered Speech (UCLASS) [15 14]. The database consists of recording for monologs, readings and conversation. There are 40 different speakers contributing 107 reading recording in the database. In this work speech samples are taken from standard reading of 25 different speakers with age between 10 years to 20 years. The samples were chosen to cover wide range of age and stuttering rate. The repetition, prolongation and filled pause dysfluencies are segmented manually by hearing the speech signal. The segmented samples are subjected to feature extraction. The same standard English passages that were used in UCLASS database are used in preparing the fluent database. Twenty fluent speakers with mean age group of 25 were made to read the passage and recorded using cool edit version 2.1.

## 3. METHODOLOGY

The overall process of dysfluent and fluent speech classification is divided into 4 steps as shown in figure 1.

## 3.1. Pre-emphasis

This step is performed to enhance the accuracy and efficiency of the feature extraction processes. This will compensate the high-frequency part that was suppressed during the sound production mechanism of humans. The speech signal *s(n)* is sent to the high-pass filter:

$$s_2(n) = s(n) - a*s(n-1) \tag{1}$$

Where $s_2(n)$ is output signal and the recommended value of *a* is usually between 0.9 and 1.0[10]. The z- transform of the filter is

$$H(z) = 1 - a*z^{-1} \tag{2}$$

The aim of this stage is to boost the amount of energy in the high frequencies.

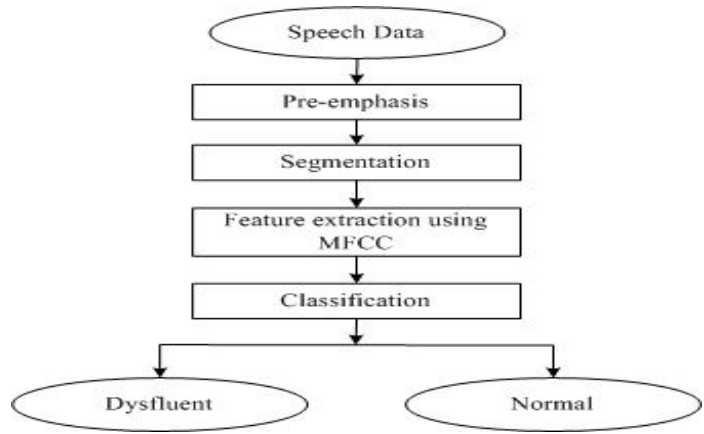

Figure 1. Schematic diagram of classification method

## 3.2. Segmentation

In this paper we are considering 3 types of dysfluencies in stuttered speech such as repetitions, prolongations and interjections; these were identified by hearing the recorded speech samples and were segmented manually. The segmented samples are subjected to feature extraction.

## 3.3. Feature Extraction (MFCC)

Feature extraction is to convert an observed speech signal to some type of parametric representation for further investigation and processing. Several feature extraction algorithms are used for this task such as Linear Predictive Coefficients (LPC), Linear Predictive Cepstral Coefficients (LPCC), Mel Frequency Cepstral Coefficients (MFCC) and Perceptual Linear Prediction (PLP) cepstra.

The MFCC feature extraction is one of the best known and most commonly used features for speech recognition. It produces a multi dimensional feature vector for every frame of speech. In this study we have considered 12MFCCs. The method is based on human hearing perceptions which cannot perceive frequencies over 1KHz. In other words, MFCC is based on known variation of the human ear's critical bandwidth with frequency [7].The block diagram for computing MFCC is given in figure 2. The step-by-step computations of MFCC are discussed briefly in the following sections.

### 3.3.1. Step 1: Framing

In framing, we split the pre-emphasis signal into several frames, such that we are analyzing each frame in the short time instead of analyzing the entire signal at once [9]. Hamming window is applied to each frame, which will cause loss of information at the beginning and end of frames. To overcome this overlapping is applied, to reincorporate the information back into extracted feature frames. The frame length is set to 25ms and there is 10ms overlap between two adjacent frames to ensure stationary between frames.

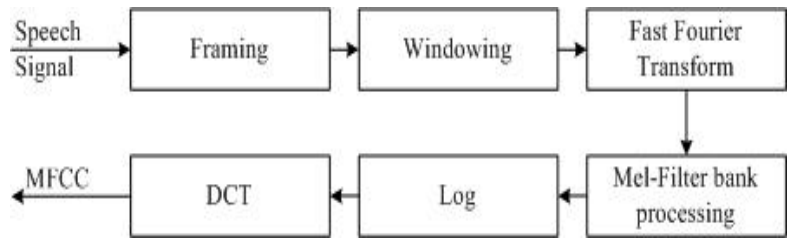

Figure 2. MFCC computation

### 3.3.2. Step 2: Windowing

The effect of the spectral artifacts from framing process is reduced by windowing [9]. Windowing is a point-wise multiplication between the framed signal and the window function. Whereas in frequency domain, the combination becomes the convolution between the short-term spectrum and the transfer function of the window. A good window function has a narrow main lobe and low side lobe levels in their transfer function [9]. The purpose of applying Hamming window is to minimize the spectral distortion and the signal discontinuities. Hamming window function is shown in following equation:

$$w(n) = 0.54 - 0.46\cos\left(\frac{2\pi n}{N-1}\right), 0 \leq n \leq N-1 \quad (3)$$

If the window is defined as $w(n)$, $0 \leq n \leq N-1$. Then the result of windowing signal is

$$Y(n) = X(n) \times W(n) \quad (4)$$

Where, $N$ = number of samples in each frame, $Y(n)$ = Output signal, $X(n)$ = input signal and $W(n)$ = Hamming window.

### 3.3.3. Step 3: Fast Fourier Transform (FFT)

The purpose of FFT is to convert the signal from time domain to frequency domain preparing to the next stage (Mel frequency wrapping). The basis of performing Fourier transform is to convert the convolution of the glottal pulse and the vocal tract impulse response in the time domain into multiplication in the frequency domain [2]. The equation is given by:

26

$$Y(w) = FFT[h(t) * X(t)] = H(w) * X(w) \tag{5}$$

If $X(w)$, $H(w)$ and $Y(w)$ are the Fourier Transform of $X(t)$, $H(t)$ and $Y(t)$ respectively.

### 3.3.4. Step 4: Mel Filter Bank Processing

A set of triangular filter banks is used to approximate the frequency resolution of the human ear. The Mel frequency scale is linear up to 1000 Hz and logarithmic thereafter [1]. A set of overlapping Mel filters are made such that their centre frequencies are equidistant on the Mel scale. The Filter banks can be implemented in both time domain and frequency domain. For the purpose of MFCC processing, filter banks are implemented in frequency domain. The filter bank according to Mel scale is shown in figure 3.

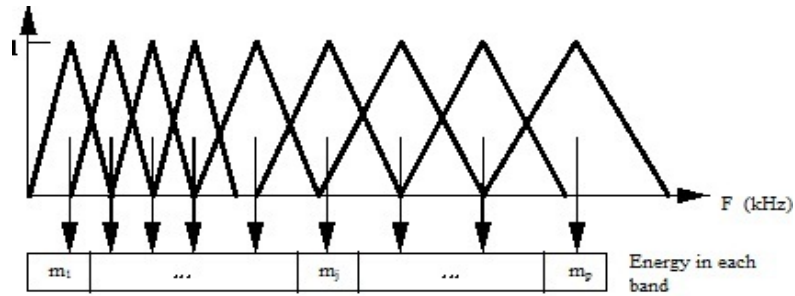

Figure 3. Mel scale filter bank

The figure 3 shows a set of triangular filters which are used to compute a weighted sum of filter spectral components and the output of the process approximates to a Mel scale. The magnitude frequency response of each filter is triangular in shape and equal to unity at the centre frequency. Also decreases linearly to zero at centre frequency of two adjacent filters. The output of each filter is sum of its filtered spectral components. Afterwards approximation of Mel's for a particular frequency can be expressed using following equation:

$$mel(f) = 2595 * \log_{10}\left(1 + \frac{f}{700}\right) \tag{6}$$

### 3.3.5. Step 5: Discrete Cosine Transform (DCT)

In this step log Mel spectrum is converted back to time domain using DCT. The outcome of conversion is called MFCCs. Since the speech signal represented as a convolution between slowly varying vocal tract impulse response (filter) and quickly varying glottal pulse (source), the speech spectrum consists of the spectral envelope (low frequency) and the spectral details (high frequency). Now, we have to separate the spectral envelope and spectral details from the spectrum.

The logarithm has the effect of changing multiplication into addition. Therefore we can simply convert the multiplication of the magnitude of the Fourier transform into addition by taking the DCT of the logarithm of the magnitude spectrum.

We can calculate the Mel frequency cepstrum from the result of the last step using equation 7[13].

$$\tilde{c}_n = \sum_{k=1}^{K} (\log \tilde{s}_k) \cos\left[n\left(k-\frac{1}{2}\right)\frac{\pi}{K}\right], n = 1, 2, 3, ...K \tag{7}$$

Where $\tilde{c}_n$ is $MFCC_i$, $\tilde{s}_k$ is Mel spectrum and $K$ is the number of cepstrum coefficients.

## 3.4. Classification

The k-Nearest Neighbor (k-NN) and SVM are used as classification techniques in the proposed approach.

### 3.4.1. k - Nearest Neighbor (k-NN)

k-NN classifies new instance query based on closest training examples in the feature space. k-NN is a type of instance-based learning, or lazy learning where the function is only approximated locally and all computation is delayed until the classification is done. Each query object (test speech signal) is compared with each of training object (training speech signal). Then the object is classified by a majority vote of its neighbors with the object being assigned to the class most common amongst its k nearest neighbors (k is a positive integer, typically small). If k = 1, then the object is simply assigned to the class of its nearest neighbor [8].

In this study minimum distance is calculated from test speech signal to each of the training speech signal in the training set. This classifies test speech sample belonging to the same class as the most similar or nearest sample point in the training set of data. A Euclidean distance measure is used to find the closeness between each training set data and test data. The Euclidean distance measure equation is given by:

$$d_e(a,b) = \sqrt{\sum_{i=1}^{n}(b_i - a_i)^2} \tag{8}$$

Our aim is to perform two class classification (dysfluent vs. fluent) using the MFCC features. We have considered two different training data set; one for dysfluent speech samples that includes 3 types of dysfluencies such as repetitions, prolongations and interjections, second training data set is for fluent speech. For each test samples the training data set is found with k nearest members. Further, for this k nearest members, suitable class label is identified based on majority voting. Class labels can be dysfluent speech or fluent speech.

### 3.4.2 Support Vector Machines (SVM)

A SVM is a classification technique based on the statistical learning theory [17, 18]. It is supervised learning technique that uses a labelled data set for training and tries to find a decision function that classifies best the training data. The purpose of the algorithm is to find a hyperplane to define decision boundaries separating between data points of different classes. SVM classifier finds the optimal hyperplane that Correctly Separates (classifies) the largest fraction of data points while maximizing the distance of either class from the hyperplane. The hyper plane equation is given by

$$w^T x + b \tag{9}$$

where $w$ is weight vector and b is bias.

Given the training labelled data set $\{x_i, y_i\}_{i=1}^{N}$ with $x_i \in \mathbb{R}^d$ being the input vector and $y_i \in \{-1, +1\}$. Where $x_i$ is input vector and $y_i$ is its corresponding label [19]. SVMs map the $d$ - dimensional input vector $x$ from the input space to the $d_h$ - dimensional feature space by non-linear function $\varphi(\cdot): \mathbb{R}^d \to \mathbb{R}_h^d$ Hence hyperplane equation becomes

$$w^T \varphi(x) + b = 0 \tag{10}$$

With $b \in \mathbb{R}$ and $w$ an unknown vector with the same dimension as $\varphi(x)$. The resulting optimization problem for SVM, is written as

$$\min_{w,\xi,b} J_1(w, \xi) = \frac{1}{2} w^T w + c \sum_{i=1}^{n} \xi_i \tag{11}$$

such that

$$y_i(w^T \varphi(x_i) + b) \geq 1 - \xi_i, \quad i = 1, \ldots, N \tag{12}$$

$$\xi_i \geq 0, \quad i = 1, \ldots, N \tag{13}$$

The constrained optimization problem in equation 11, 12 and 13 is referred as the primal optimization problem. The optimization problem of SVM is usually written in dual space by introducing restriction in the minimizing functional using Lagrange multipliers. The dual formulation of the problem is

$$\max_{\alpha} \sum_{i=1}^{m} \alpha_i - \frac{1}{2} \sum_{i,j=1}^{N} \alpha_i \alpha_j y_i y_j (x_i, x_j) \tag{14}$$

subject to $\alpha_i \geq 0$ for all $i = 1, \ldots m$ and $\sum_{i=1}^{m} \alpha_i y_i = 0$

Thus, the hyperplane can be written in the dual optimization problem as:

$$f(x) = sgn\left[\sum_{i=1}^{m} y_i \alpha_i (x_i, x) + b\right] \tag{15}$$

## 4. RESULTS AND DISCUSSIONS

The samples were chosen as explained in section 2 of this paper. The database is divided into two subsets: training set and testing set based on the ratio 80:20 respectively. The Table 2 shows the distribution of speech segments for training and testing. To analyze speech samples first we extract MFCC feature, afterwards two training database is constructed for dysfluent and fluent speech samples. Once the system is trained, test set is employed to estimate the performance of classifiers.

Table 2. The speech data

|  | Speech samples | Training | Testing |
|---|---|---|---|
| Dysfluent speech | 50 | 40 | 10 |
| Fluent speech | 50 | 40 | 10 |

The experiment was repeated 3 times, each time different training and testing sets were built randomly. The result of training and testing for dysfluent and fluent speech is shown in Table 3. Figure 4 shows the average classification result.

Table 3. Dysfluent and fluent classification result with 3 different set

| Data set | k-*NN* | | *SVM* | |
|---|---|---|---|---|
| | **Dysfluent** | **Fluent** | **Dysfluent** | **Fluent** |
| *Set 1* | 80 | 90 | 90 | 90 |
| *Set 2* | 90 | 90 | 90 | 100 |
| *Set 3* | 90 | 100 | 90 | 100 |
| *Average Classification (%)* | 86.67 | 93.34 | 90 | 96.67 |

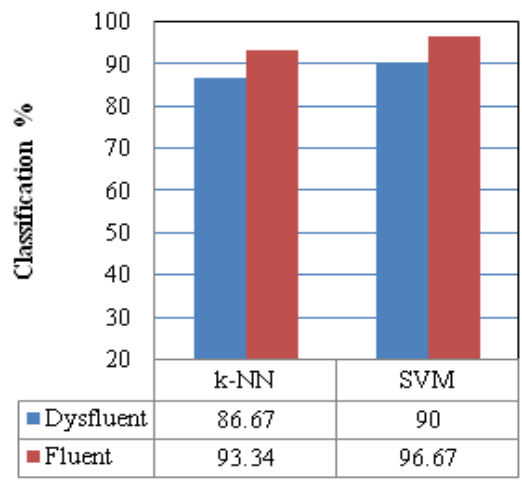

Figure 4. Average classification results of k-NN and SVM classifiers

## 5. CONCLUSIONS

The speech signal can be used as a reliable indicator of speech abnormalities. We have proposed an approach to discriminate dysfluent and fluent speech based on MFCC feature analysis. Two classifiers such as *k*-NN and SVM were applied on MFCC feature set to classify dysfluent and

International Journal of Computer Science, Engineering and Applications (IJCSEA) Vol.2, No.6, December 2012

fluent speech. Using *k*-NN classifier we have obtained an average accuracy of 86.67% and 93.34% for dysfluent and fluent speech respectively. The SVM classifier yielded an accuracy of 90% and 96.67% for dysfluent and fluent speech respectively. In this work we have considered combination of three types of dysfluencies which are important in classification of dysfluent speech. In the future work number of training data can be increased to improve the accuracy of testing data and different feature extraction algorithm can be used to improve the performance.

**Authors**

**P. Mahesha** received his Bachelor's Degree in Electronics and Communications Engineering from University of Mysore, Karnataka, India. Master's Degree in Software Engineering from the Visvesvaraya Technological University (VTU), Belgaum, Karnataka, India and currently he is pursuing PhD under VTU. He has published 4 International Conference papers related to his research area. He is currently working as Assistant Professor at the Department of Computer Science and Engineering, Sri Jayachamarajendra College of Engineering, Mysore, Karnataka, India. He has 7 years of teaching experience. His research interests include Speech Signal Processing, Web Technologies and Software Engineering. 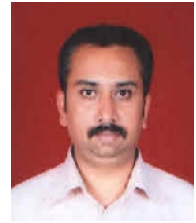

**D.S. Vinod** received his Bachelor's Degree in Electronics and Communications Engineering and Master's Degree in Computer Engineering from the University of Mysore, Karnataka, India. He has completed PhD at Visvesvaraya Technological University (VTU), Belagaum, Karnataka, India. He did his research work on Multispectral Image Analysis and published 2 International Journals and 10 International Conference papers related to his research area. He is currently working as Assistant Professor at the Department of Information Science and Engineering, Sri Jayachamarajendra College of Engineering, Mysore, Karnataka, India. He has 13 years of teaching experience and he was awarded UGC-DAAD Short-term fellowship, Germany in the year 2004-05. His research interests include Image Processing, Speech Signal Processing, Machine Learning and Algorithms. 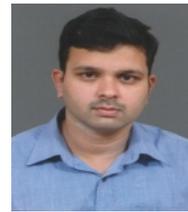